\documentstyle[epsf]{res}
\begin{document}

\title{ Clustering statistics on a light-cone in the cosmological
 redshift space}

\author{Yasushi SUTO \\
{\it Department of Physics and Research Center for the Early Universe\\
School of Science, University of Tokyo, Tokyo 113-0033, Japan, \\
suto@phys.s.u-tokyo.ac.jp}
}

\maketitle

\section*{Abstract}

We summarize a series of our recent work concerning the cosmological
redshift-space distortion and light-cone effects.  After briefly
describing the theoretical formalism, we show how those effects are
sensitive to the cosmological parameters. Then we apply this formalism
to predict the two-point correlation functions and power spectra for
the X-ray clusters, galaxies and quasars in future surveys, and
discuss their cosmological implications.


\section{Cosmological effects in the high-z universe}

Redshift surveys of galaxies definitely serve as the central database
for observational cosmology. In addition to the existing catalogues
including CfA1, CfA2, SSRS, and the Las Campanas survey, upcoming
galaxy surveys such as 2dF (2-degree Field Survey) and SDSS (Sloan
Digital Sky Survey) are expected to provide important clues to the
evolution of structure in the universe.  In addition to those {\it
  shallower} surveys, clustering in the universe in the range $z= 1 -
3$ has been partially revealed by, for instance, the Lyman-break
galaxies and X-ray selected AGNs. In particular, the 2dF and SDSS QSO
redshift surveys promise to extend the observable scale of the
universe by an order of magnitude, up to a few Gpc. A proper
interpretation of such redshift surveys in terms of the clustering
evolution, however, requires an understanding of many cosmological
effects which can be neglected for $z\ll 1$ and thus have not been
considered seriously so far.  These cosmological {\it contaminations}
include linear redshift-space (velocity) distortion (Kaiser 1987),
nonlinear redshift-space (velocity) distortion (e.g., Suto \&
Suginohara 1991; Cole, Fisher, \& Weinberg 1994), cosmological
redshift-space (geometrical) distortion (Alcock \& Paczy\'nski 1979;
Ballinger, Peacock, \& Heavens 1996; Matsubara \& Suto 1996), and
cosmological light-cone effect (Yamamoto \& Suto 1999; Suto et al.
1999; Yamamoto, Nishioka \& Suto 1999).

We describe a theoretical formalism to incorporate those effects, in
particular the cosmological redshift-distortion and light-cone
effects, and present several specific predictions in cold dark matter
(CDM) models.

\section{ Cosmological redshift-space distortion}

Due to a general-relativistic effect through the geometry of the
universe, the {\it observable} separations perpendicular and parallel
to the line-of-sight direction, $x_{{\rm s}{\scriptscriptstyle\perp}} =
(c/H_0)z\delta\theta$ and $x_{{\rm s}{\scriptscriptstyle\parallel}}=
(c/H_0)\delta z$, are mapped differently to the corresponding comoving
separations in real space $x_{{\scriptscriptstyle\perp}}$ and
$x_{{\scriptscriptstyle\parallel}}$:
\begin{eqnarray}
\label{eq:x2xs}
x_{{\rm s}{\scriptscriptstyle\perp}} (z) &=& x_{\perp} cz/[H_0
  (1+z)d_{\rm\scriptscriptstyle {A}}(z)] \equiv x_{\perp}/c_\perp(z),
\\
 x_{{\rm s}{\scriptscriptstyle\parallel}} (z) &=& x_{\parallel}
  H(z)/H_0 \equiv x_{\parallel}/c_\parallel(z) ,
\end{eqnarray}
with $d_{\rm\scriptscriptstyle {A}}(z)$ being the angular diameter
distance. The difference between $c_\perp(z)$ and $c_\parallel(z)$
generates an apparent anisotropy in the clustering statistics, which
should be isotropic in the comoving space. Then the power spectrum in
cosmological redshift space, $P^{({\rm\scriptscriptstyle {CRD}} )} $,
is related to $P^{({\rm S})}$ defined in the {\it comoving} redshift
space as
\begin{equation}
\label{eq:crdrel}
  P^{({\rm\scriptscriptstyle {CRD}} )}
(k_{{\rm s}\perp},k_{{\rm s}\parallel};z) 
  =\frac{1}{c_\perp(z)^2c_\parallel(z)}
P^{({\rm S})} \left(\frac{k_{{\rm s}{\scriptscriptstyle\perp}}}{c_\perp(z)},
\frac{k_{{\rm s}{\scriptscriptstyle\parallel}}}{c_\parallel(z)};z \right) ,
\end{equation}
where the first factor comes from the Jacobian of the volume element
$dk_{{\rm s}{\scriptscriptstyle\perp}}^2 dk_{{\rm
    s}{\scriptscriptstyle\parallel}}$, and $k_{{\rm s}\perp}=
c_\perp(z) k_{\perp}$ and $k_{{\rm s}\parallel}= c_\parallel(z)
k_{\parallel}$ are the wavenumber perpendicular and parallel to the
line-of-sight direction.  If one assumes a scale-independent
deterministic linear bias, the power spectrum distorted by the
peculiar velocity field, $P^{({\rm S})}(k;z)$, is known to be well
approximated by the following expression (Cole et al. 1995; Peacock \&
Dodds 1996):
\begin{equation}
\label{eq:power_in_redshiftspace}
  P^{({\rm S})}(k_\perp,k_\parallel;z) 
  = b^2(z)P^{({\rm R})}_{{\rm\scriptscriptstyle mass}}(k;z)
\left[1+\beta(z) \left(\frac{k_\parallel}{k}\right)^2 \right]^2
D\left[k_\parallel{\hbox {$\sigma_{\scriptscriptstyle {\rm P}}$}}(z)\right],
\end{equation}
where $k_\perp$ and $k_\parallel$ are the comoving wavenumber
perpendicular and parallel to the line-of-sight of an observer, and
$P^{({\rm R})}_{{\rm\scriptscriptstyle mass}}(k;z)$ is the mass power
spectrum in real space. The finger-of-god effect is modeled by the
damping function, $D\left[k_\parallel{\hbox
    {$\sigma_{\scriptscriptstyle {\rm P}}$}}(z)\right]$, for which we
assume a Lorentzian. Then equation (\ref{eq:crdrel}) reduces to
\begin{eqnarray}
\label{eq:powercrd}
P^{({\rm\scriptscriptstyle {CRD}} )}(k_{\rm s},\mu_k;z)
  &=&\frac{b^2(z)}{c_\perp(z)^2c_\parallel(z)}
P^{({\rm R})}_{\rm\scriptscriptstyle mass} \left(\frac{k_{\rm s}}{c_\perp(z)}
  \sqrt{1+\left[{1\over \eta(z)^2}-1\right]\mu_k^2} ; z \right) \cr &&
  \hspace*{-4cm} \times \left\{1+ \left[{1\over
  \eta(z)^2}-1\right]\mu_k^2 \right\}^{-2} \left\{1+ \left[{1+\beta(z)
  \over \eta(z)^2}-1\right]\mu_k^2\right\}^2 ~ \left[1+
  \frac{k_{\rm s}^2\mu_k^2{\hbox {$\sigma_{\scriptscriptstyle {\rm
  P}}$}}^2}{2c^2_\parallel(z)} \right]^{-1},
\end{eqnarray}
where we introduce
\begin{eqnarray}
\label{eq:k2ks}
k_{\rm s} \equiv \sqrt{ k_{{\rm s}\perp}^2 + k_{{\rm s}\parallel}^2}, \quad
  \mu_k \equiv k_{{\rm s}\parallel}/k_{\rm s}, \quad
  \eta \equiv c_\parallel/c_\perp,
\end{eqnarray}
following Ballinger et al. (1996) and Magira et al. (2000).

\begin{figure}[t]
\begin{center}
 \leavevmode\epsfxsize=12.0cm \epsfbox{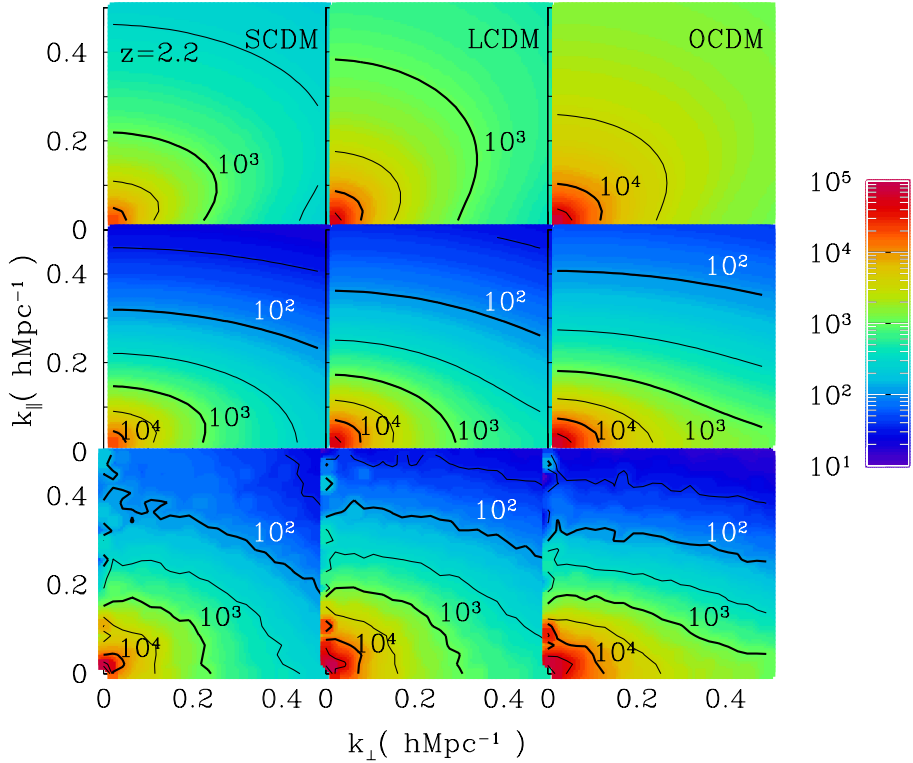}
\caption{
  Two-dimensional power spectra in cosmological redshift space at
  $z=2.2$.  \label{fig:power2d}}
\end{center}
\end{figure}
Figure \ref{fig:power2d} shows anisotropic power spectra
$P^{({\rm\scriptscriptstyle {CRD}} )}(k_{\rm s},\mu_k;z=2.2)$.  As
specific examples, we consider SCDM (standard CDM), LCDM (Lambda CDM),
and OCDM (Open CDM) models, which have $(\Omega_0, \lambda_0, h,
\sigma_8)$ $= (1.0, 0.0, 0.5, 0.6)$, $(0.3, 0.7, 0.7, 1.0)$, and
$(0.3, 0.0, 0.7, 1.0)$, respectively.  These sets of cosmological
parameters are chosen so as to reproduce the observed cluster
abundance (Kitayama \& Suto 1997).  Our theoretical predictions use
the fitting formulae of Peacock \& Dodds (1996; PD) for the nonlinear
power spectrum, $P^{({\rm R})}_{{\rm\scriptscriptstyle {mass}} }
(k;z)= 2\pi^2\Delta^2_{\rm\scriptscriptstyle NL}(k,z)/k^3 $, of Mo,
Jing, \& B\"{o}rner (1997) for the pair-wise peculiar velocity
dispersions:
\begin{eqnarray}
  \sigma_{\scriptscriptstyle{\rm\scriptscriptstyle P,MJB}}^2 &\equiv& 
    \Omega(z)H_0^2
   \left[1-\frac{1+z}{D_+^2(z)}
   \int_z^\infty \frac{D_+^2(z')}{(1+z')^2}dz'\right] 
   \int_0^{\infty}\frac{dk}{k}
 \frac{\Delta^2_{\rm\scriptscriptstyle NL}(k,z)}{k^2} ,
\end{eqnarray}
with $D_+(z)$ being the linear growth rate.  Clearly the linear theory
predictions ($\sigma_{\scriptscriptstyle {\rm P}}=0$; top panels) are
quite different from the results of N-body simulations (bottom
panels), indicating the importance of the nonlinear velocity effects
($\sigma_{\scriptscriptstyle {\rm
    P}}=\sigma_{\scriptscriptstyle{\rm\scriptscriptstyle P,MJB}}$;
middle panels).

\begin{figure}[t]
\begin{center}
 \leavevmode\epsfxsize=12.0cm \epsfbox{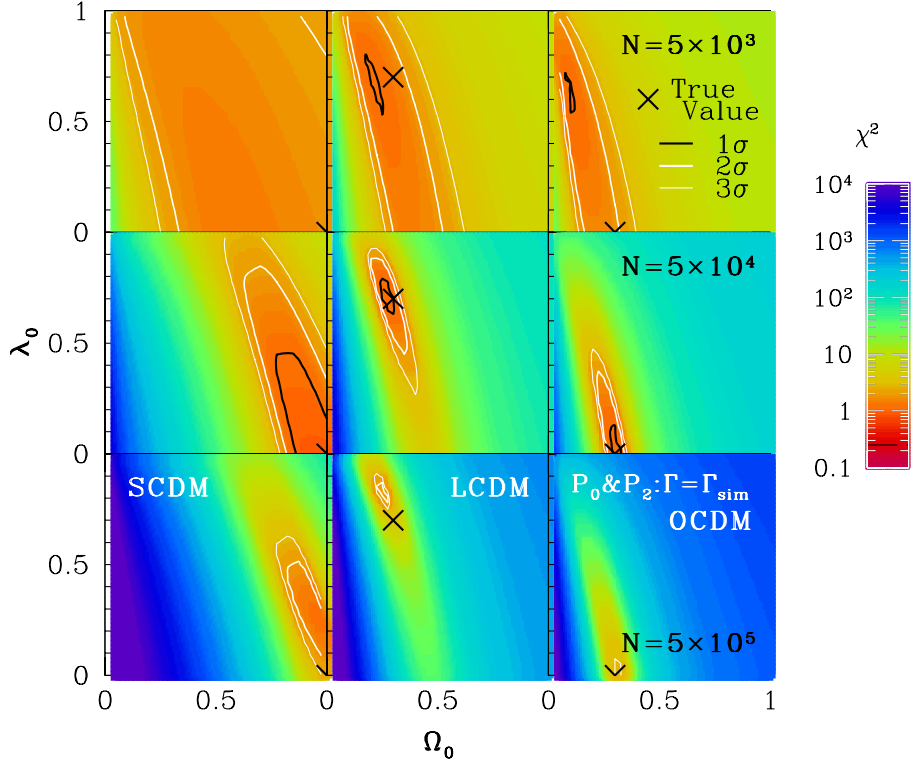}
\caption{The confidence contours on  $\Omega_0$-$\lambda_0$  plane
  from the $\chi^2$-analysis of the monopole and quadrupole moments of
  the power spectrum in the cosmological redshift space at $z=2.2$.
  We randomly selected $N=5\times10^3$ (upper panels), $N=5\times10^4$
  (middle panels), and $N=5\times10^5$ (lower panels) particles from
  N-body simulation. The value of $\sigma_8$ is adopted from the
  cluster abundance.
\label{fig:chi2p02ol}}
\end{center}
\end{figure}
Next we decompose the power spectrum into harmonics:
\begin{eqnarray}
\label{eq:pkmoment}
P(k,\mu_k;z) = \sum_{l: {\rm even}} P_l(k) L_l(\mu_k), \quad 
P_l(k;z) \equiv 
\frac{2l+1}{2}\int^1_{-1}d\mu_k P(k,\mu_k;z) L_l(\mu_k) ,
\end{eqnarray}
where $L_l(\mu_k)$ are the $l$-th order Legendre polynomials.
Similarly, the two-point correlation function is decomposed as
\begin{eqnarray}
\label{eq:ximoment}
\xi(x,\mu_x;z) = \sum_{l: {\rm even}} \xi_l(x) L_l(\mu_x), \quad 
\xi_l(x;z) \equiv 
\frac{2l+1}{2}\int^1_{-1}d\mu_x \xi(x,\mu_x;z) L_l(\mu_x) ,
\end{eqnarray}
using the direction cosine, $\mu_x$, between the separation vector and
the line-of-sight.  The above multipole moments satisfy the following
relations:
\begin{eqnarray}
\label{eq:pk2xi}
\xi_l(x;z) = {1 \over 2\pi^2 i^l} \int_0^\infty P_l(k;z) j_l(kx) k^2dk,
\end{eqnarray}
with $j_l(kx)$ being spherical Bessel functions. Substituting
$P^{({\rm\scriptscriptstyle {CRD}} )}(k_{\rm s},\mu_k;z)$ in equation
(\ref{eq:pkmoment}) yields $P^{({\rm\scriptscriptstyle {CRD}} )}_
l(k_{\rm s};z)$, and then $\xi^{({\rm\scriptscriptstyle {CRD}} )}({\bf
  x_s};z)$ can be computed from equation (\ref{eq:pk2xi}).

Comparison of the monopoles and quadrupoles from simulations and model
predictions exhibits how the results are sensitive to the cosmological
parameters, which in turn may put potentially useful constraints on
$(\Omega_0, \lambda_0)$.  Figure \ref{fig:chi2p02ol} indicates the
feasibility, which interestingly results in a constraint fairly
orthogonal to that from the Supernovae Ia Hubble diagram.

\section{Cosmological light-cone effect}

Observing a distant patch of the universe is equivalent to observing
the past.  Due to the finite light velocity, a line-of-sight direction
of a redshift survey is along the time, as well as spatial, coordinate
axis. Therefore the entire sample does not consist of objects on a
constant-time hypersurface, but rather on a light-cone, i.e., a null
hypersurface defined by observers at $z=0$. This implies that many
properties of the objects change across the depth of the survey
volume, including the mean density, the amplitude of spatial
clustering of dark matter, the bias of luminous objects with respect
to mass, and the intrinsic evolution of the absolute magnitude and
spectral energy distribution. These aspects should be properly taken
into account in order to extract cosmological information from
observed samples of redshift surveys. We apply the formulation on the
light-cone originally developed by Yamamoto \& Suto (1999) to X-ray
selected clusters and on-going SDSS galaxy and QSO catalogues.

\subsection{ Two-point correlation functions of 
  X-ray selected clusters}

Provided an X-ray flux-limited sample of clusters ($S>S_{\rm lim}$),
it is fairly straightforward to compute its two-point correlation
function $\xi^{\rm S}_{\rm cl}(R,z;>S_{\rm lim})$ at a given $z$; a
fairly accurate empirical expression for the bias parameter $b(z)$ as
a function of the halo mass is known (e.g., Jing 1998), and the mass
is translated to the X-ray temperature assuming the virial
equilibrium, and then to the X-ray luminosity from the observed
luminosity-temperature relation (e.g., Kitayama \& Suto 1996).  The
corresponding correlation function {\it on the light-cone} is given by
\begin{figure}[t]
\begin{center}
 \leavevmode\epsfxsize=11.5cm \epsfbox{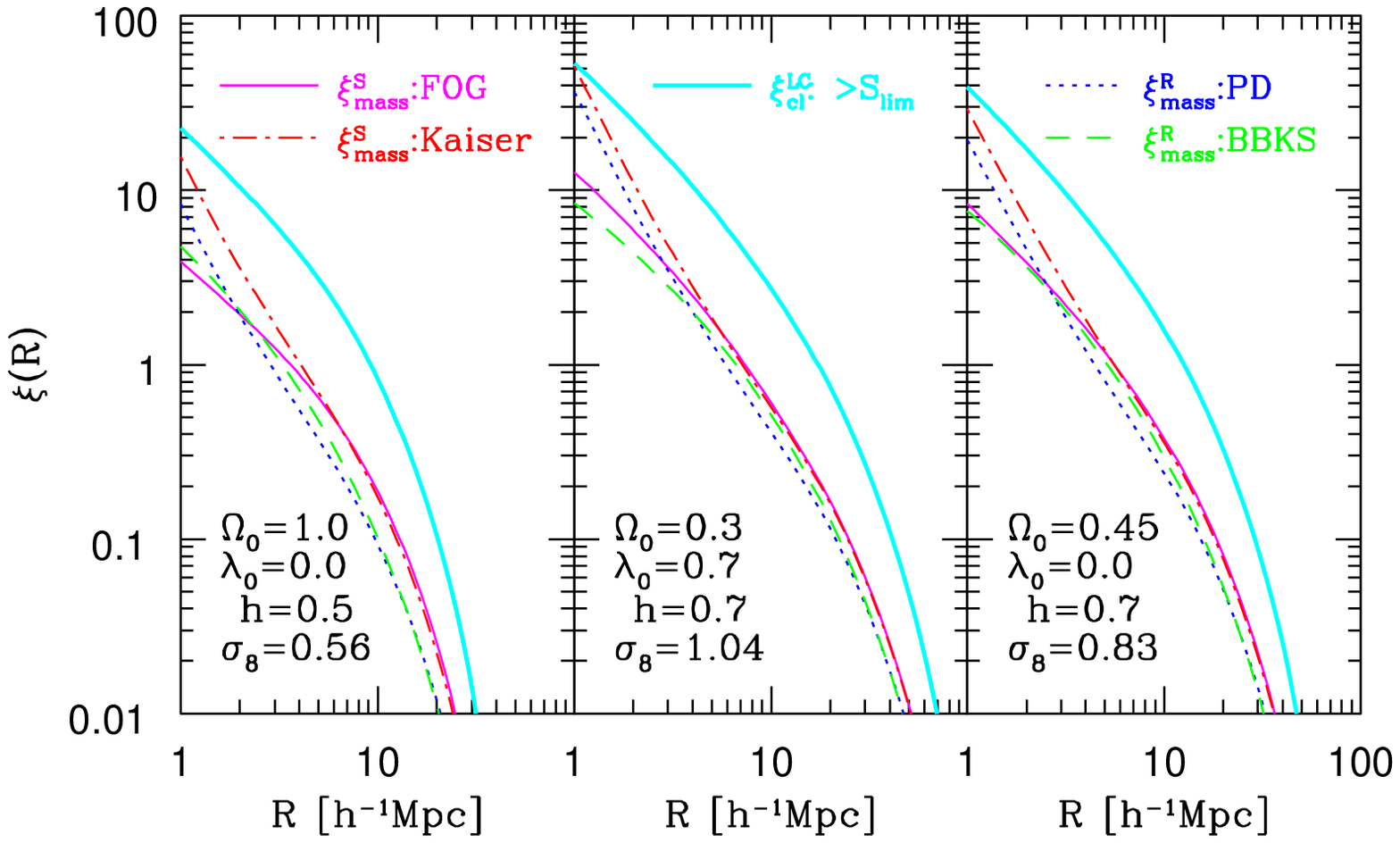}
\caption{Light-cone and redshift-space distortion effects on
  two-point correlation functions of clusters.
\label{fig:xicl_diff}
}
\end{center}
\end{figure}
\begin{eqnarray}
\label{eq:lcxir1}
    \xi^{\rm LC}_{\rm X-cl}(R; >S_{\rm lim}) 
= {
   {\displaystyle 
     \int_{z_{\rm min}}^{z_{\rm max}} dz 
     {dV_{\rm c} \over dz} ~n_0^2(z)
    \xi^{\rm S}_{\rm cl}(R,z(r);>S_{\rm lim})
    }
\over
    {\displaystyle
     \int_{z_{\rm min}}^{z_{\rm max}} dz 
     {dV_{\rm c} \over dz} ~n_0^2(z)
     }
} ,
\end{eqnarray}
where $R$ is the comoving separation of a pair of clusters, $z_{\rm
  max}$ and $z_{\rm min}$ denote the redshift range of the survey, and
$dV_{\rm c}/dz$ is the comoving volume element per unit solid angle
(Suto et al. 2000; Moscardini et al. 2000).  The comoving number
density of clusters in the flux-limited survey, $n_0(z; >S_{\rm
  lim})$, is computed by integrating the Press -- Schechter mass
function.

\begin{figure}[t]
\begin{center}
 \leavevmode\epsfxsize=11.0cm \epsfbox{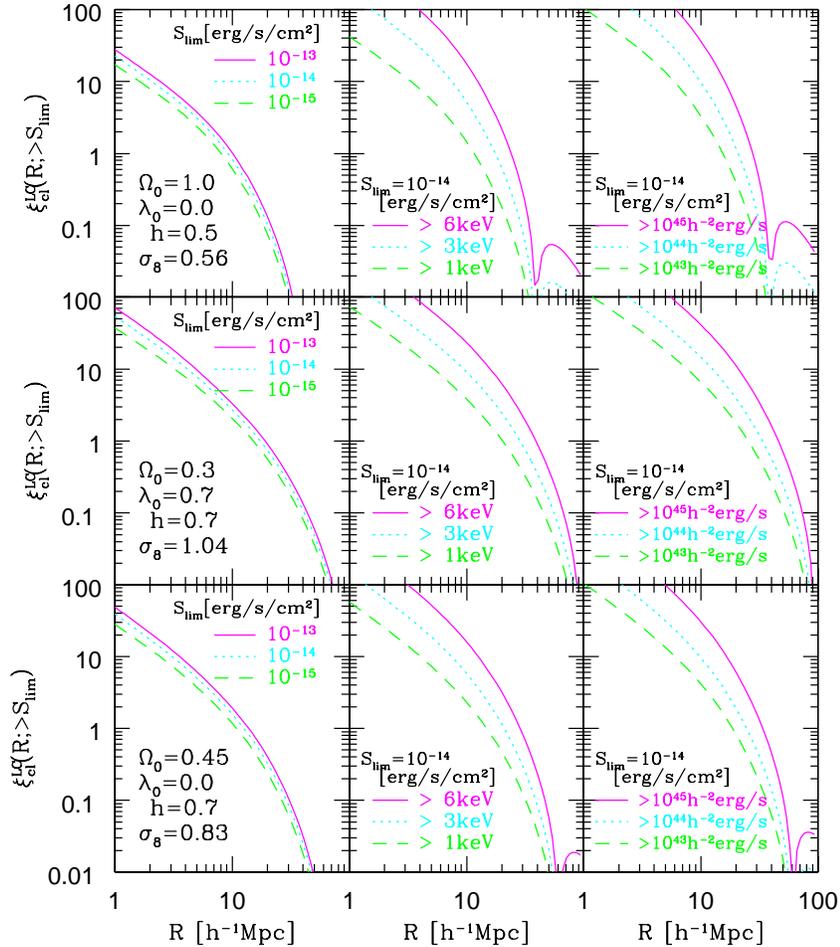}
\caption{ Two-point correlation functions of clusters in SCDM ({\it
    Top panels}), LCDM ({\it Middle panels}), and OCDM ({\it Bottom
    panels}) for different selection criteria.
  \label{fig:xicl_sxtl}}
\end{center}
\end{figure}

\begin{figure}[t]
\begin{center}
 \leavevmode\epsfxsize=7.0cm \epsfbox{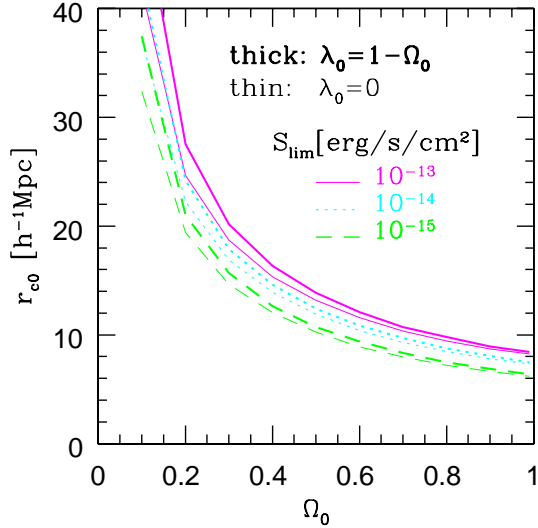}
\caption{Correlation lengths of clusters as a function of
  $\Omega_0$. The shape parameter of the spectrum $\Gamma$ is fixed as
  $\Omega_0 h \exp[-\Omega_{\rm b}(1+\sqrt{2 h}\Omega_0^{-1})]$ with
  $\Omega_{\rm b}h^2=0.015$ and $h=0.7$.  
\label{fig:rc0_omega0}} 
\end{center}
\end{figure}

Figure \ref{fig:xicl_diff} plots several predictions for two-point
correlation functions under different assumptions; linear and
nonlinear mass correlations in real space at $z=0$ using the Bardeen
et al. (1986; BBKS) and PD formulae for mass power spectra, cluster
correlations with linear redshift-space distortion (Kaiser 1987) and
with full redshift-space distortion at $z=0$ using
$\sigma_{\scriptscriptstyle{\rm\scriptscriptstyle P,MJB}}$. These
should be compared with our final predictions on the light-cone in
redshift space (with $S_{\rm lim}=10^{-14}$erg/s/cm$^2$; thick solid
lines).  Figure \ref{fig:xicl_sxtl} shows our predictions for
$\xi^{\rm LC}_{\ cl}(R)$ for cluster samples selected with different
flux-limit $S_{\rm lim}$ ({\it left panels}), and with additional
temperature and absolute bolometric luminosity limits, $T_{\rm lim}$
and $L_{\rm lim}$ ({\it middle and right panels}).  For the latter two
cases, $S_{\rm lim}=10^{-14}$erg/s/cm$^2$ is assumed for definiteness.
The results are insensitive to $S_{\rm lim}$, but very sensitive to
$T_{\rm lim}$ and $L_{\rm lim}$, reflecting the strong dependence of
the bias on the latter quantities.
For a cosmological application of the present result, it is
interesting to examine how $r_{c,0}(S_{\rm lim})$ defined through
\begin{eqnarray}
\label{eq:rc0}
\xi^{\rm LC}_{\rm cl}(r_{c0};>S_{\rm lim})=1
\end{eqnarray}
depends on $\Omega_0$. This is summarized in Figure
\ref{fig:rc0_omega0}, where we fix the value of the fluctuation
amplitude $\sigma_8$ adopting the cluster abundance constraint
(Kitayama \& Suto 1997).  Again the results are not sensitive to the
flux limit $S_{\rm lim}$.  The dependence on $\Omega_0$ is rather
strong, and these predictions combined with the future observational
results will be able to break the degeneracy of the cosmological
parameters.

\subsection{Power spectra of SDSS galaxy and QSO samples}

\begin{figure}[t]
\begin{center}
 \leavevmode\epsfxsize=10.0cm \epsfbox{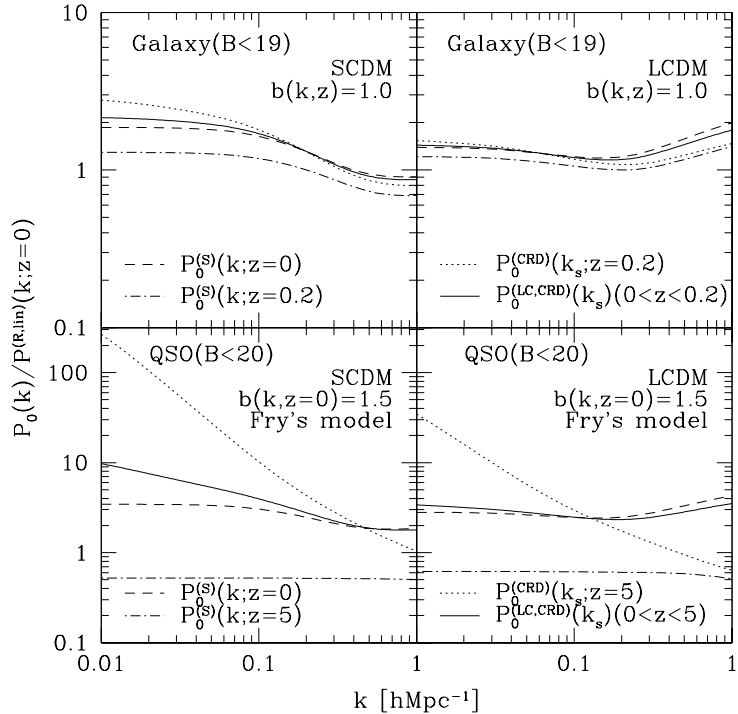}
\caption{Light-cone and cosmological redshift-space 
  distortion effects on angle-averaged power spectra. 
\label{fig:pk_lccrd}}
\end{center}
\end{figure}

Finally we present theoretical predictions of power spectra relevant
for SDSS galaxy and QSO samples, fully taking account of the
cosmological redshift-space distortion and light-cone effects.
Denoting the comoving number density and the selection function of the
objects by $n_0^{{\rm\scriptscriptstyle {com}}}(z)$, and $\phi(z)$,
Suto, Magira \& Yamamoto (2000) obtain
\begin{eqnarray}
\label{eq:lccrdpkmom}
    P^{({\rm\scriptscriptstyle {LC}} 
,{\rm\scriptscriptstyle {CRD}} )}_l(k_{\rm s}) 
&=& {
   {\displaystyle 
     \int_{z_{\rm min}}^{z_{\rm max}} dz 
     {dV_{\rm c} \over dz} ~[\phi(z)n_0^{{\rm\scriptscriptstyle {com}} }(z)]^2
    {c_\perp(z)^2c_\parallel(z)} P_l^{({\rm\scriptscriptstyle {CRD}} )}(k_{\rm s};z)
    }
\over
    {\displaystyle
     \int_{z_{\rm min}}^{z_{\rm max}} dz 
     {dV_{\rm c} \over dz}  ~[\phi(z)n_0^{{\rm\scriptscriptstyle {com}} }(z)]^2
          {c_\perp(z)^2c_\parallel(z)}
     }
} .
\end{eqnarray}
Figure \ref{fig:pk_lccrd} compares several predictions for the
angle-averaged (monopole) power spectra normalized by the real-space
counterpart in linear theory.  The upper and lower panels adopt the
selection functions appropriate for galaxies in $0<z<{z_{\rm
    max}}=0.2$ and QSOs in $0<z<{z_{\rm max}}=5$, respectively. The
left and right panels present the results in SCDM and LCDM models. For
simplicity we adopt a scale-independent linear bias model of Fry
(1996),
\begin{equation} 
  b(z)= 1 +{1\over D_+(z)} [b(k,z=0)-1],
\label{FryM}
\end{equation}
with $b(k,z=0)=1$ and $1.5$ for galaxies and quasars, respectively.
It is clear that the cosmological redshift-space distortion and the
light-cone effect substantially change the predicted shape and
amplitude of the power spectra, even for the SDSS galaxy sample.

\section{ Summary and conclusions}

We have presented a theoretical formalism to predict the two-point
clustering statistics on a light-cone in the cosmological redshift
space.  The present methodology will find two completely different
applications.  For relatively shallower catalogues like galaxy
samples, the evolution of bias is not supposed to be so strong. Thus,
one may estimate the cosmological parameters from the observed degree
of the redshift distortion, as has been conducted conventionally. Most
importantly, one can now correct for the systematics due to the
light-cone and geometrical distortion effects, which affect the
estimate of the parameters by $\sim 10$\%.  Alternatively, for deeper
catalogues like high-redshift quasar samples, one can extract
information on the nonlinearity, scale-dependence and stochasticity of
the object-dependent bias only by correcting the observed data on the
basis of our formulae. In this case, although one should adopt a set
of cosmological parameters a priori, those will be provided both from
the low-redshift analysis described above and from precision data of
the cosmic microwave background and supernovae Ia.  In a sense, the
former approach uses the light-cone and geometrical distortion effects
as real cosmological signals, while the latter regards them as
inevitable, but physically removable, noise. In both cases, the
present methodology is essential in properly interpreting the
observations of the universe at high redshifts.

\vspace{2mm} 

I thank Y.P.Jing, Tetsu Kitayama, Hiromitsu Magira, Takahiko
Matsubara, Hiroaki Nishioka, and Kazuhiro Yamamoto for enjoyable
collaborations on which the present talk is based.  Numerical
computations were carried out on VPP300/16R and VX/4R at the
Astronomical Data Analysis Center of the National Astronomical
Observatory, Japan, as well as at RESCEU and KEK (National Laboratory
for High Energy Physics, Japan).  This research was supported in part
by the Grants-in-Aid by the Ministry of Education, Science, Sports and
Culture of Japan to RESCEU (07CE2002).

\section*{References}
\baselineskip=14pt
\re
1. \  Alcock C., Paczy\'nski B.\ 1979, Nature 281, 358
\re
2. \ Ballinger W.E., Peacock J.A., Heavens A.F.\ 1996, MNRAS 282, 877
\re
3. \ Bardeen J.M., Bond J.R., Kaiser N., Szalay A.S.\ 1986, ApJ 304, 15
\re
4. \   Cole S., Fisher K.B., Weinberg D.H.\ 1994, MNRAS 267, 785
\re
5. \  Fry J.N.\ 1996, ApJ 461, L65
\re
6. \  Jing Y.P.\ 1998, ApJ 503, L9
\re
7. \  Kaiser N.\ 1987, MNRAS 227, 1
\re
8. \  Kitayama T., Suto Y.\ 1996, ApJ 469, 480
\re
9. \   Kitayama T., Suto Y.\ 1997, ApJ 490, 557
\re
10.\   Magira H., Jing Y.P., Suto Y.\ 2000, ApJ 528, 30 
\re
11.\   Matsubara T., Suto Y.\ 1996, ApJ 470, L1
\re
12.\  Mo H.J., Jing Y.P., B\"orner G.\ 1997, MNRAS 286, 979 
\re
13.\ Moscardini, L., Matarrese, S., Lucchin, F., \& Rosati, P.\ 2000;
   MNRAS, submitted (astro-ph/9909273);
\re
14.\  Peacock J.A., Dodds S.J.\ 1996, MNRAS 280, L19
\re
15.\  Suto Y., Magira H., Jing Y.P., Matsubara T.,  Yamamoto K.\ 1999,
  Prog.\ Theor.\ Phys.\ Suppl. 133, 183
\re
16.\ Suto Y., Magira H., Yamamoto K.\ 2000, PASJ, 52, No.2, in press
\re
17.\ Suto Y., Suginohara T.\ 1991, ApJ 370, L15
\re
18.\ Suto Y.,  Yamamoto K., Kitayama T., Jing Y.P.\ 2000, ApJ 
534, May 10 issue, in press
\re
19.\ Yamamoto K., Nishioka H., Suto Y.\ 1999, ApJ 527, 488
\re 
20.\ Yamamoto K.,  Suto Y.\ 1999, ApJ 517, 1

\end{document}